\documentclass[a4paper]{jpconf}
\usepackage{graphicx}
\usepackage{dsfont}
\begin{document}
\title{Coherent~center~domains~from~local~Polyakov~loops}

\author{S.~Borsanyi$^{\,1}$, \underline{J.~Danzer}$^{\;2,\dagger}$, 
Z.~Fodor$^{\,1}$, 
C.~Gattringer$^{\,2}$, A.~Schmidt$^{\,2}$}


\address{$^1$ University of Wuppertal, Department of Physics, 
42097 Wuppertal, Germany}
\address{$^2$ Karl-Franzens University Graz, Institute of Physics, 
8010 Graz, Austria}
\address{$^\dagger$ Speaker. E-mail: {\tt julia.danzer@uni-graz.at}}

\begin{abstract}
We analyze properties of local Polyakov loops using quenched as well as
dynamical SU(3) gauge configurations for a wide range of temperatures. It is
demonstrated that for both, the confined and the deconfined regime, the local
Polyakov loop prefers phase values near the center elements $1,
e^{\pm i 2\pi/3}$. We divide the lattice sites into three sectors
according to these phases and show that the sectors give rise to the
formation of clusters. For a suitable definition of these clusters we find
that in the quenched case deconfinement manifests itself as the onset of
percolation of the clusters. A possible continuum limit of the
center clusters is discussed.
\end{abstract}

\section{Motivation and general framework}

\vskip2mm

With the running and upcoming experiments at LHC, RHIC and GSI the analysis
of the QCD phase diagram has become an important focus of research. Not only
the transition  curves which separate different phases are of interest, but
one would also like to understand  from first principles the physical
mechanisms that drive the various transitions.

In this project we probe the finite temperature transition of QCD using
static quark sources. In the framework of lattice QCD these can be
implemented using local Polyakov loops. The local Polyakov loop $L(\vec{x})$
is given by the trace of the product of temporal gauge links $U_4(\vec{x},t)$
at a fixed spatial position $\vec{x}$ ($N_t$ is the number of lattice points
in time direction): 
\begin{equation}
L(\vec{x}) \; = \; \mbox{Tr} \; \prod_{t=0}^{N_t-1}\ U_4(\vec{x},t) \;  ,
\label{PloopL}
\end{equation}
i.e., the Polyakov loop is a gauge transporter that propagates a static quark
at position $\vec{x}$ forward in time. For later use we also define
the spatially averaged Polyakov loop $P$  as $P =
V^{-1}\sum_{\vec{x}}L(\vec{x})$, where $V$ is the spatial volume. Due to
translation invariance $P$ and $L(\vec{x})$ have the same vacuum expectation
value. After suitable renormalization the Polyakov loop may be related to the
free energy $F_q$ of a single quark, i.e.,  $\langle P \rangle \propto
\exp{(-F_q/T)}$, where $T$ is the temperature. Below $T_c$ the free energy is
infinite, thus $\langle P \rangle=0$, and the quarks are confined. Above
$T_c$ we have a finite free energy and thus $\langle P \rangle\ne 0$,
signaling deconfinement. Hence the Polyakov loop acts as an order parameter
for deconfinement.

In the deconfined phase of pure SU(3) gauge theory the phases of the summed Polyakov loops $P$ assume 
values in the vicinity of the three center elements $1, e^{\pm i 2\pi/3}$ of
SU(3). This behavior shows the underlying center symmetry, a symmetry of the
action and the path integral measure, which in pure gauge theory becomes
broken spontaneously above $T_c$. The Polyakov loop $P$ transforms
non-trivially under the center transformation and thus is also an order parameter
for the symmetry breaking. Above $T_c$ the Polyakov loop $\langle P \rangle$
is non-vanishing with phases near $1, e^{\pm i 2\pi/3}$. For full QCD
the fermion determinant breaks the center symmetry explicitly and acts as
an external magnetic field favoring the real sector. 

Except for the interchange of low and high temperature the situation is equivalent
to simple spin systems with spins $s(\vec{x})$.  Without
external field the corresponding Hamiltonian   has a symmetry which
may be broken spontaneously. The symmetry breaking can be analyzed with
observables that transform non-trivially under the symmetry group, e.g., the
magnetization $M = V^{-1} \sum_{\vec{x}} s(\vec{x})$.  To obtain the
equivalence between the spin system and the gauge theory one has to
identify the local loops $L(\vec{x})$ with the spins $s(\vec{x})$ and the
spatially averaged Polyakov loop $P$ with the magnetization $M$. In the
absence of an external field (the fermion determinant) the phase of the
magnetization $\langle M \rangle$ (the spatially averaged Polyakov loop
$\langle P \rangle$) spontaneously selects one of the phases according to 
the underlying symmetry. If the external field (the fermion determinant)
is turned on, the previously sharp transition becomes a crossover, and the
phase of the magnetization (the Polyakov loop) is determined by the 
symmetry breaking term.

For the case of pure gauge theory these arguments are the basis of the
Svetitsky-Yaffe conjecture \cite{SYC} which states that at $T_c$ 
pure SU(3) gauge theory in 4 space-time dimensions can be described by a 3-d
effective spin system with an effective action which is symmetric under the
center group $\mathds{Z}_3$. The spin degrees of freedom are related to the
local loops $L(\vec{x})$. The leading term of the effective action is given
by ($\tau$ and $\kappa$ are real non-negative couplings)
\begin{equation}
S[s] \, = \, - \tau \sum_{\langle x,y \rangle}
\Big[ s(\vec{x})s(\vec{y})^* \, + \, s(\vec{y})s(\vec{x})^* \Big] 
\; - \; 
\kappa \sum_{\vec{x}} \Big[ s(\vec{x}) + s(\vec{x})^* \Big] \; ,
\label{effact}
\end{equation}
where for illustration purposes we also included a symmetry breaking term
which can be turned off when $\kappa = 0$. In the simplest version the
effective spins $s(\vec{x})$ have values $s(\vec{x}) \in \{1, e^{\pm i 2\pi/3}\}$.

Magnetic finite temperature transitions for spin systems are well understood
phenomena. For discrete symmetry groups the transition is accompanied by the
formation of locally spin coherent Weiss domains near $T_c$, 
before one spin orientation wins out in the symmetry broken
phase. Even if a modest external magnetic field is applied one can still
observe
local clusters of aligned spins  different from the direction preferred by
the magnetic field. 

One can go a step further and analyze the connectedness properties of the
Weiss domains. One may define clusters of aligned spins and switch to a
cluster description of magnetic systems \cite{FK,FK2,Coniglioklein}. For several
spin systems one finds that the
magnetic transition may be characterized as a percolation phenomenon of
suitably defined clusters.

With the Svetitsky-Yaffe conjecture in mind, which describes the deconfinement
transition with an effective spin system, we can formulate the central
questions we explore in our project:

\begin{itemize}

\item Can one identify (at least near $T_c$) characteristic properties of
spin-like behavior in an ab-initio lattice simulation of pure gauge theory
and/or full  QCD?  

\item Is it possible to identify spatial structures (clusters) that
correspond to Weiss domains?

\item How do the domains behave near $T_c$? Do suitably defined clusters
percolate?

\item What is the role of the fermion determinant which breaks the underlying
center symmetry?

\item Can the clusters (Weiss domains) be given a physical meaning also in
the continuum limit?

\end{itemize}

For SU(2) gauge theory similar questions were addressed in 
\cite{Fortunato1} -- \cite{Fortunatospincluster} (see also \cite{Alex}). First results
for SU(3) gauge theory were presented in \cite{Gattringer}.

\section{Distribution properties of local Polyakov loops}

\vskip2mm

We study the Polyakov loop (\ref{PloopL}) using quenched
configurations as well as configurations from full QCD. For our quenched
analysis   we use the L\"uscher-Weisz gauge action with lattice sizes from
$20^3 \times 6$ to $40^3 \times 12 $ and temperatures ranging from $T =
0.63\, T_c$ to $1.32\, T_c$ \cite{Gattringer}. In full QCD we use
configurations generated with a Symanzik improved gauge action and $2+1$
flavors of stout-link improved staggered quarks at physical masses
\cite{WB-group1,WB-group2}. We study lattices with sizes $18^3 \times 6 ,
36^3 \times 6$ and $24^3 \times 8$ in a temperature range from  $T = 110$ MeV
to $320$ MeV. 

To analyze the properties of local Polyakov loops we evaluate 
the $L(\vec{x})$ and write them as
\begin{equation}
L(\vec{x}) \; = \; \rho(\vec{x}) \, e^{i\varphi(\vec{x})} \, ,
\end{equation}
i.e., we decompose the local loops into modulus and phase. We begin with
studying histograms $H[\rho(\vec{x})]$ and $ H[\varphi(\vec{x})]$  for the
distribution of the local modulus $\rho(\vec{x})$ and the local phase
$\varphi(\vec{x})$, shown in Figs.~1 and 2. In both figures the  lhs.~is for
the quenched case, while the rhs.~is for full QCD and we compare results in
the confining (low $T$) and the deconfining (high $T$) phase.

\begin{figure}[b]
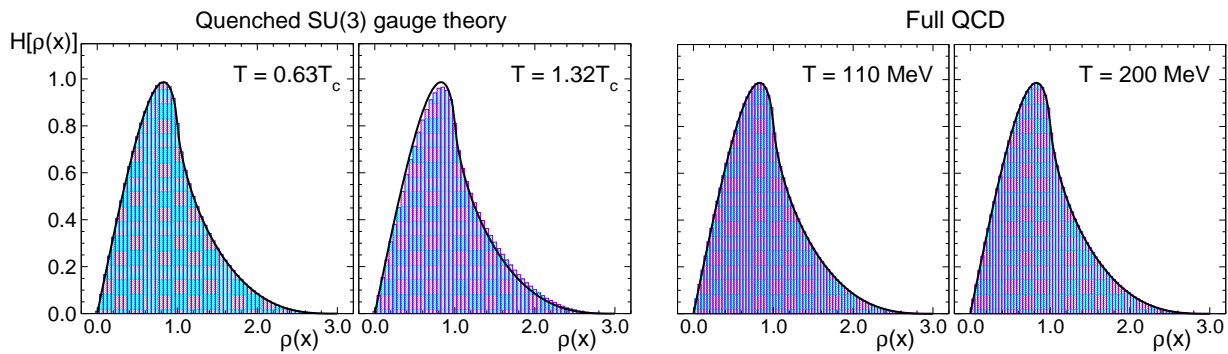

\hspace*{-1mm}
\includegraphics[height=45.5mm,clip]{Pr_histo_40x6.eps} \hspace{-1mm}
\includegraphics[height=45.5mm,clip]{histo_36x6_rad_dyn.eps}
\caption{\label{rho}
Histograms for the distribution of the local modulus  $\rho(\vec{x})$. We
compare quenched (lhs.) and full QCD (rhs.) at low  and high $T$. The full
curve is the Haar measure distribution.}
\end{figure}

\begin{figure}[t]
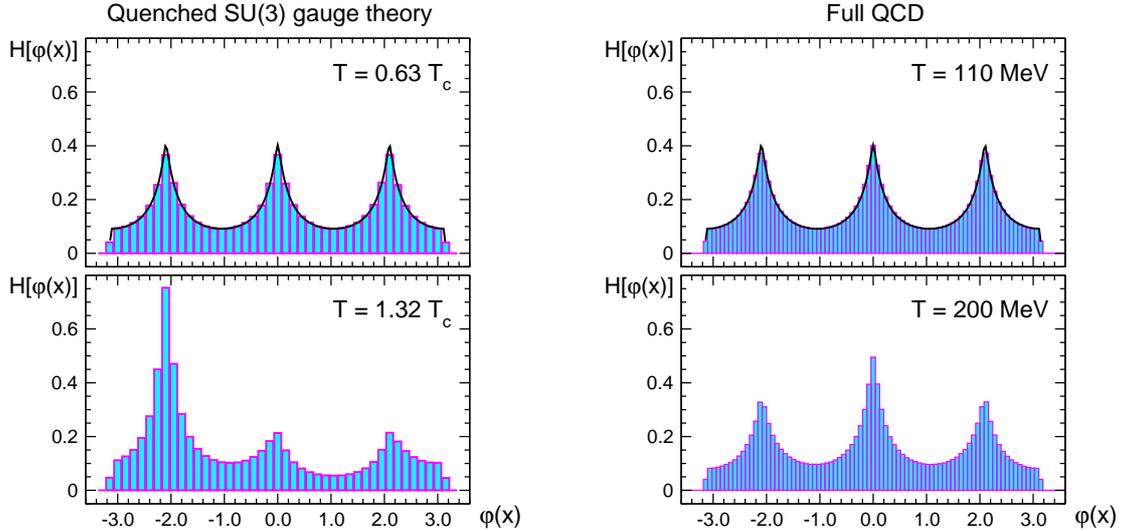

\centering
 \hspace*{-2mm}
\includegraphics[height=70mm,clip]{histo_thetaP_rot_40x6.eps} \hspace{8mm}
\includegraphics[height=70mm,clip]{histo_36x6_phase_dyn.eps}
\caption{\label{phase}
Histograms for the distribution of the local phase $\varphi(\vec{x})$. We
compare quenched (lhs.) and full QCD (rhs.) at low  and high $T$. The full
curve is the Haar measure distribution.}
\end{figure}

Fig.~1 shows that the distribution of the modulus is 
essentially independent of $T$, and furthermore is the same 
for both quenched and full QCD. The distribution of the modulus almost 
perfectly follows the Haar measure distribution (full curves in 
Fig.~1),
$P[\rho] = \int\!dU \delta(\rho -|\mbox{Tr}\, U | )$, where $dU$ is
the SU(3) Haar measure. Comparing the distributions below and above the
transition/crossover temperature clearly shows that the local modulus
is not involved in the transition, which in the case of pure SU(3) gauge 
theory is even manifest as a first order jump of $\langle P \rangle$. 
Obviously it must be the local phases $\varphi(\vec{x})$ which drive the
transition.

The histograms for the local phase in Fig.~2 show a pronounced peak 
structure, with maxima at the three center angles 0, $2\pi/3$ and $-2\pi/3$. 
In the confining phase (top plots) for all three maxima have 
the same height and we again observe no difference between the quenched and 
the dynamical case, both of which can be described by the Haar measure 
distribution $P[\varphi] = \int\!dU \delta(\varphi - \mbox{arg Tr} \, U)$
(full curves in the top plots).
In the deconfined phase (bottom plots) the situation is 
different. One of the three center phases is more populated. For full QCD,
where the fermion determinant acts like an external field, it is always the 
real center sector (phase values near 0) which becomes enhanced. In the
quenched case any of the three center sectors may be selected spontaneously
(similar to, e.g., the Ising system where the magnetization has two 
signs to choose from in the symmetry broken phase). In the quenched 
distribution we show here we have chosen configurations where the phase angle
of the summed Polyakov loop $P$ is near $- 2\pi/3$ and obviously the local 
phases are enhanced for that value. If one switches to any of the other equivalent
center sectors, the distribution is shifted by $\pm 2\pi/3$.

\section{Cluster properties}

\vskip2mm

In the previous section we have seen that the transition to the deconfined  phase is
accompanied by an increasing population of the histograms for the phase
$\varphi(\vec{x})$ of the local Polyakov loop $L(\vec{x})$ near one of the center
angles. This accumulation of the local phases in one sector drives the increase
of the expectation value $\langle P \rangle$, while the modulus of  the local loops
plays no role. It is interesting to note that at high temperatures (bottom plots
in Fig.~2) there are still pronounced peaks also in the subdominant sectors. The
question is whether the phase values at different positions $\vec{x}$ are
distributed independently, or if there  are spatial domains where the phases of
the local Polyakov loops tend to align in the same center sector. The latter
case is what is suggested by the effective action (\ref{effact}), where the first
term favors parallel spins. 

In order to study the formation of domains we assign sector numbers 
$n(\vec{x})$ to the sites $\vec{x}$,
\begin{equation}
n(\vec{x}) \; = \; \left\{ \begin{array}{rl}
-1 & \; \mbox{for} \;\;\;\; \varphi(\vec{x}) \, \in \, 
[\,-\pi + \delta \; , \; -\pi/3 - \delta \, ] \; ,\\
0 & \; \mbox{for} \;\;\;\; \varphi(\vec{x}) \, \in \, 
[\,-\pi/3 + \delta \, , \, \pi/3 - \delta \, ] \; ,\\
+1 & \; \mbox{for} \;\;\;\; \varphi(\vec{x}) \, \in \, 
[\,\pi/3 + \delta \, , \, \pi - \delta \,] \; . 
\end{array} \right. 
\end{equation}
Here $\delta$ is a free real and positive parameter which allows to cut
lattice points $\vec{x}$ where the corresponding phase $\varphi(\vec{x})$ 
is near a minimum of the distributions in Fig.~2. These points do not 
have a clear preference for one of the center sectors and the parameter 
$\delta$ allows one to remove them from the cluster analysis. 
The remaining lattice points $\vec{x}$ can now be organized in clusters according to 
the sector numbers: We put two neighboring points $\vec{x}, \vec{y}$ 
into the same cluster if $n(\vec{x}) = n(\vec{y})$. For illustration purposes
in Fig.~3 we show the largest cluster for two quenched configurations, one below 
$T_c$ (lhs.), the other one above (rhs.). The plot is for lattice size 
$30^3 \times 6$ and a value of $\delta$ chosen such that 39 \% of the
lattices points are cut. It is obvious, that above $T_c$ 
the largest cluster percolates (stretches over all of the lattice), 
while it is finite below $T_c$. 

Of course the cutoff parameter $\delta$ will influence the size of the 
clusters, since with increasing $\delta$ less sites are available. 
We stress at this point that also for the characterization of 
the magnetic transitions in spin systems a similar reduction step is 
necessary in the construction of the clusters. This may be a reduced linking
probability between neighbors with equal spins 
\cite{Coniglioklein}, but also more general approaches similar 
to the one used here were considered \cite{Fortunatospincluster}. As a consistency check 
of our cluster construction one may show \cite{inprep} that the points that survive the 
cut carry most of the signal of a rising Polyakov loop above 
the transition/crossover temperature.

\begin{figure}[t]
\begin{center}
\includegraphics[width=70mm,clip]{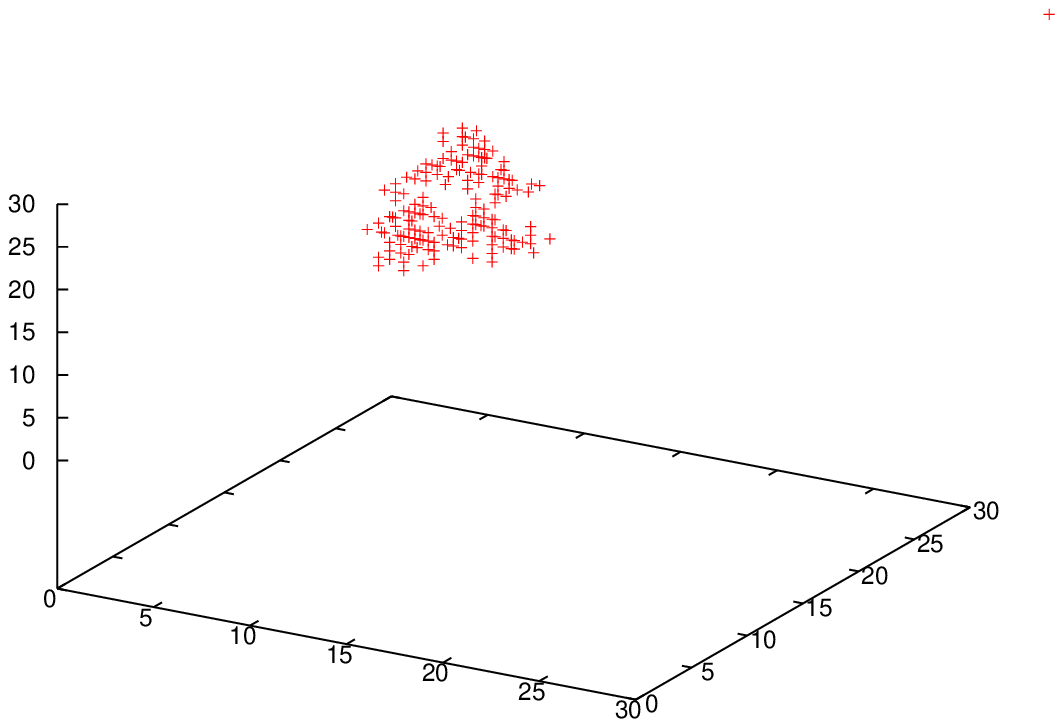} 
\hskip3mm
\includegraphics[width=70mm,clip]{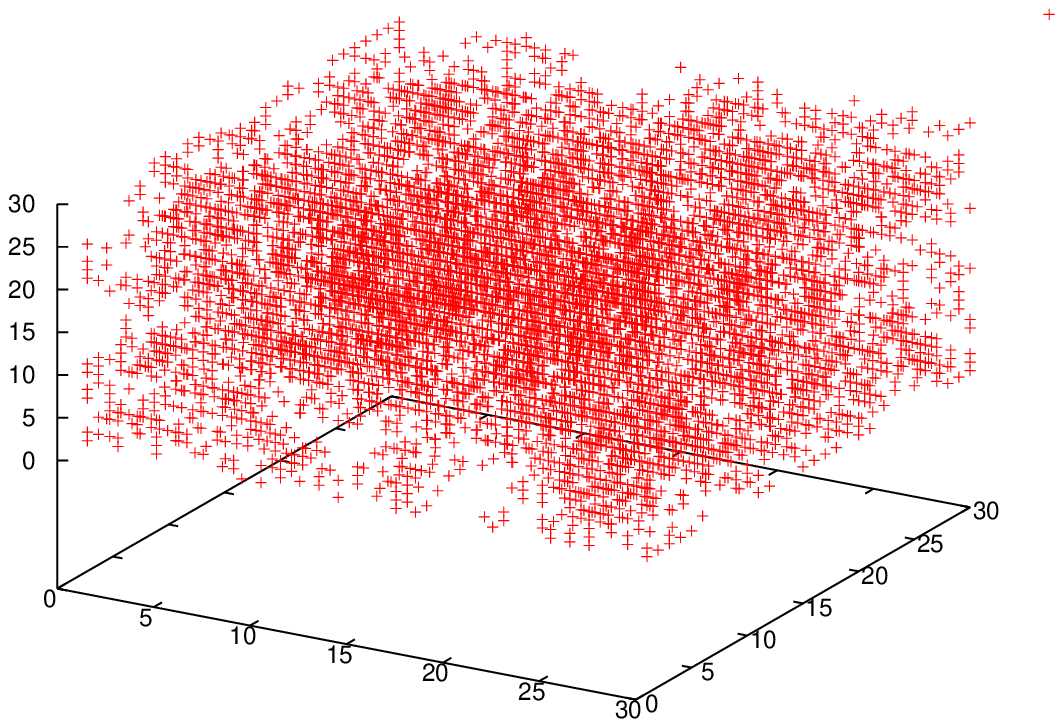} 
\end{center}
\caption{\label{cluster3D}
Largest clusters for two quenched configurations below $T_c$ (lhs.) and above 
(rhs.).}
\end{figure}

\begin{figure}[b]
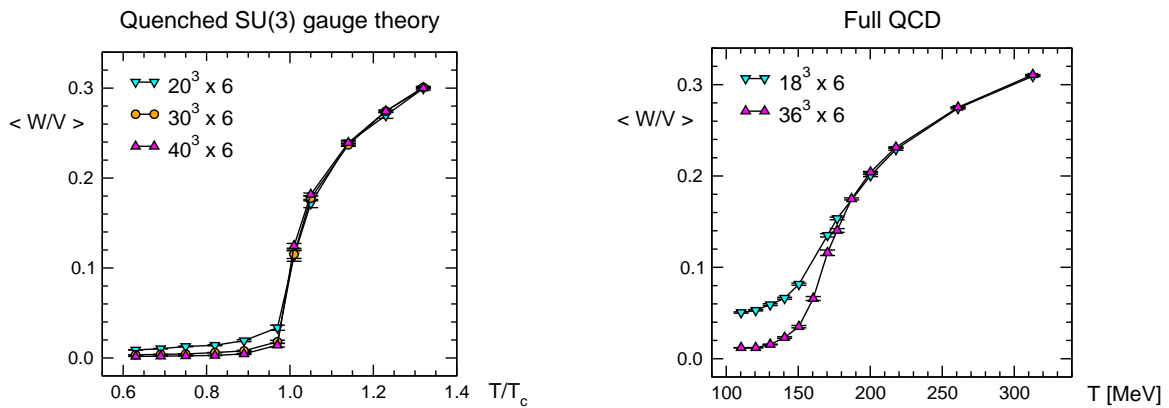

\centering
\includegraphics[height=53mm,clip]{maxcluster_finsize.eps}
\hskip10mm
\includegraphics[height=53mm,clip]{maxcluster_finsize_dyn.eps}
\caption{\label{largest}
Weight $W$ of the largest cluster normalized with the volume $V$ as function of $T$.}
\end{figure}

In order to quantify the dependence of the cluster size on the temperature,
in Fig.~4 we show the expectation value of the weight $W$ 
(i.e., the number of lattice points) of the 
largest cluster normalized by the total number $V$ of sites as a function of
temperature. Again the lhs.\ is for the quenched case (at a cut of 39\% for
all volumes and temperatures), 
while the rhs.\ is 
for full QCD (19 \% cut), and we compare three (two) 
different spatial volumes.
Both in the quenched and the dynamical case we observe 
that the clusters start out small below $T_c$ and grow quickly in 
size at the transition temperature. If one analyzes the percolation 
probability one finds that for the quenched case, where we have a true
phase transition, this probability indeed approaches a step function as
the volume is increased. For the dynamical case, where one observes only a 
crossover \cite{crossover}, 
the situation concerning percolation is not yet clear and further 
studies on several volumes will be necessary to settle this issue.
 
\section{Towards a continuum limit for the center clusters}

\vskip2mm

We have shown that the phases $\varphi(\vec{x})$ 
of the local Polyakov loops $L(\vec{x})$
have preferred values near the center angles, and that neighboring sites 
have a tendency towards aligning these phases. The corresponding clusters
were found to grow quickly near the transition/crossover temperature, and 
at least for the quenched case the deconfinement transition may be 
characterized by the onset of percolation for suitably defined clusters. 
So far this picture is established only for a fixed lattice 
spacing $a$ and some arbitrarily chosen value of the cutoff parameter 
$\delta$ which enters our cluster definition. If one wants to assign a 
physical significance to the center clusters, a way of constructing them 
such that a continuum limit can be taken is necessary. 
In particular this involves
a prescription for connecting the scale $a$ in physical units to the clusters.

We begin with defining the linear extension (radius) of a cluster by  considering  the
expectation value of two-point correlators $C(|\vec{x} - \vec{y}|)$  for sites $\vec{x},
\vec{y}$ within the same  cluster. These 2-point functions decay exponentially,
$C(|\vec{x} - \vec{y}|) \propto \exp(- |\vec{x} - \vec{y}|/r)$, and we use the parameter
$r$ to define the radius of the cluster in units of the  lattice spacing $a$. Using the
value of the lattice constant $a$ in fm,  the diameter of the clusters in physical units
(fm) is then given by  $d_{phys} = ra$. The result will depend on both, the lattice
spacing $a$ and  the cutoff $\delta$. In order to compare the physical size of the
clusters  for different lattice resolutions $a$, we always adjust $\delta$  such that at
a low temperature ($T = 0.63 \, T_c$ for the quenched case where  this analysis is done)
we fix the physical diameter to a typical hadronic  size, e.g., $d_{phys} = 0.5$ fm. The
corresponding value of $\delta$ is then  kept fixed for all other temperatures, and is
used to study  $d_{phys}$ as a function of $T$.   In Fig.~5 we show the result for the
cluster diameter $d_{phys}$ in physical units as  a function of temperature for the
quenched case, comparing two different resolution  scales $a$ (i.e., $N_t = 6$ and $N_t =
8$). It is obvious that also in physical units the phase transition is  signaled by a
sudden increase of the cluster size. Furthermore, the results for the two different
scales fall onto a universal curve, which suggests that  a continuum limit for the
cluster size might exist. This question is  analyzed in detail in a future publication
\cite{inprep}, where we  study the flow of the cutoff $\delta$ as a function of the
resolution scale  $a$, arguing that the center clusters indeed have a continuum limit. 

\vskip1mm

\noindent
{\bf Acknowledgments:} We thank Christian Lang and Axel Maas for interesting 
discussions. This work was partly supported by the DFG SFB TR 55 and the 
FWF DK 1203.  

\begin{figure}[t]
\centering
\includegraphics[width=97mm,clip]{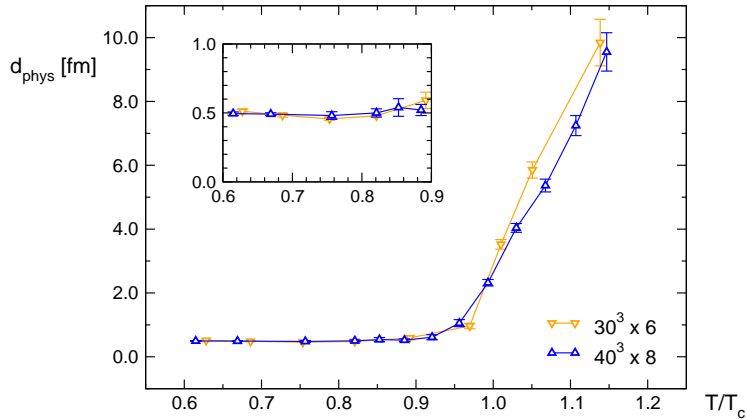}
\caption{\label{diameterfm}
Average cluster diameter in fm versus $T$ for two resolution scales  
(quenched case).}
\end{figure}

\end{document}